\documentclass[aip,jcp,amsmath,preprint,reprint]{revtex4-1}
\usepackage{amsmath}
\usepackage{dcolumn}
\usepackage{hyperref}
\usepackage{graphicx}

\DeclareMathOperator{\erf}{erf}
\DeclareMathOperator{\erfc}{erfc}
\DeclareMathOperator{\asin}{asin}
\DeclareMathOperator{\asinh}{asinh}

\renewcommand{\vec}[1]{{\rm\bf #1}}
\newcommand{\frct}[2]{{\textstyle\frac{#1}{#2}}}
\newcommand{\al}{\alpha}
\newcommand{\be}{\beta}
\newcommand{\ga}{\gamma}
\newcommand{\ka}{\kappa}
\newcommand{\la}{\lambda}
\newcommand{\om}{\omega}
\newcommand{\de}{\,\mathrm{d}}

\begin{document}

\title{Simple exchange hole models for long-range-corrected density functionals}

\author{Dimitri N. Laikov}
\email[E-Mail: ]{laikov@rad.chem.msu.ru}
\homepage[Homepage: ]{http://rad.chem.msu.ru/~laikov/}
\affiliation{Chemistry Department, Moscow State University,
119991 Moscow, Russia}

\date{\today}

\begin{abstract}
Density functionals with a range-separated treatment of the exchange energy
are known to improve upon their semilocal forerunners and fixed-fraction hybrids.
The conversion of a given semilocal functional into its short-range analog
is not straightforward, however, and not even unique,
because the latter has a higher information content that has to be recovered in some way.
Simple models of the spherically-averaged exchange hole
as an interpolation between the uniform electron gas limit
and a few-term Hermite function are developed here
for use with generalized-gradient approximations,
so that the energy density of the error-function-weighted Coulomb interaction
is given by explicit closed-form expressions
in terms of elementary and error functions.
For comparison, some new non-oscillatory models in the spirit of earlier works
are also built and studied,
their energy densities match rather closely (within less than 5\%)
but do lack the exact uniform electron gas limit.
\end{abstract}

\maketitle

It is the generalized-gradient approximation~\cite{LM83,B86,PW86,PBE96} that paved the way
for the density functional theory~\cite{HK64,KS65} into the mysterious kingdom
of theoretical chemistry. Even more fruitful may seem to be the hybrids~\cite{B93,B93b,PEB96}
with a fixed fraction of exact exchange, they are widely used,
but their ``strange'' asymptotic behavior of the effective potential is more than an \ae{}sthetic problem.
Luckily, a wonderful solution~\cite{ITYH01} was found by splitting the two-electron interaction
within the exchange energy into the short- and long-range parts
and using a density-functional approximation for the former and the full exact exchange for the latter
(the general idea has a longer history~\cite{GAP96,LSWS97}).
This was soon shown~\cite{TTYYH04} to be even more helpful
to the time-dependent~\cite{RG84,PGG96} density functional theory
where it greatly improves the calculated excited state properties
and overcomes the failure for charge-transfer excitations~\cite{DWH03}.

Given a well-tested semilocal density functional for exchange,
it is not straightforward to get its short-range analog
because the latter has a higher information content
that cannot be recovered uniquely.
The earliest studies~\cite{ITYH01,TTYYH04} took a somewhat simplistic shortcut
that breaks the underlying sum rules,
while a consistent construction should be based on an explicit model of the exchange hole ---
an entity deeply rooted in the adiabatic-connection approach~\cite{H84}.
An elegant analytic model~\cite{EP98} designed around a non-oscillatory (nodeless)
approximation~\cite{PW92} for the uniform electron gas was the first to be used~\cite{HSE03}
in this role, but the lack of closed-form expressions for the needed integrals
led to a further work~\cite{HJS08} where a similar but more computationally tractable
nodeless function has been built and proved successful both in applications~\cite{RMH09,WHS09}
and as a starting point for more sophisticated developments~\cite{TBS17}.
Other models are known, those based on an atomic-like exchange hole~\cite{BR89} are reported~\cite{MHCS16}
that satisfy fewer exact constraints, as well as oscillatory models~\cite{TM16,PJS18}
based on a density matrix expansion~\cite{NV72,KOS96,TH00}.

As we wanted to use the long-range corrected functionals for the good of chemistry,
we could not blindly adopt any such model,
we did not like the need for fitting a function to the numerical solution
of a parametrized nonlinear equation~\cite{HJS08},
we were also slightly worried about the lack of the exact uniform electron gas limit
by any nodeless exchange hole model.
We have found new and simpler explicit solutions in closed form
that should work no less well and are easy to deal with.
 
A generalized-gradient approximation for the exchange energy has a simple functional form
\begin{equation}
E_\mathrm{x} = c_\mathrm{x} \int \rho^{4/3}(\vec{r}) f\bigl(s(\vec{r})\bigr) \de^3 \vec{r} ,
\end{equation}
\begin{equation}
c_\mathrm{x} = -\frct34 \sqrt[3]{\frct{3}{\pi}} ,
\end{equation}
with all its wisdom condensed in the enhancement factor $f(s)$,
a function of only one dimensionless variable
\begin{equation}
s(\vec{r}) = \frac{\bigl|\nabla\rho(\vec{r})\bigr|}{2 p(\vec{r}) \rho(\vec{r})} ,
\end{equation}
\begin{equation}
p(\vec{r}) = \sqrt[3]{3\pi^2 \rho(\vec{r})} .
\end{equation}
On the other hand, the exact exchange energy
\begin{equation}
E_\mathrm{x} = \frct12 \int \rho(\vec{r})
\int \frac{\rho_\mathrm{x}\left(\vec{r},\vec{r}+\vec{R}\right)}{|\vec{R}|}
\de^3 \vec{R} \de^3 \vec{r}
\end{equation}
can be given in terms of the exchange hole $\rho_\mathrm{x}(\vec{r}_1,\vec{r}_2)$
whose spherically-averaged part is only needed and is then approximated
\begin{equation}
\rho_\mathrm{x}(\vec{r},\vec{r}+\vec{R}) =
\rho(\vec{r})\, q\bigl(p(\vec{r})\, |\vec{R}|, s(\vec{r}) \bigr)
\end{equation}
using the shape function $q(r,s)$
that holds more information than is otherwise hidden, by the integration,
behind $f(s)$.
If the shape function is known, the error-function-weighted short-range part of the exchange energy
\begin{equation}
E_\mathrm{x}^\om = \frct12 \int \rho(\vec{r})
\int \frac{\rho_\mathrm{x}\left(\vec{r},\vec{r}+\vec{R}\right)}{|\vec{R}|}
\erfc\bigl(\om |\vec{R}|\bigr) \de^3 \vec{R} \de^3 \vec{r}
\end{equation}
can be cast in the form
\begin{equation}
E_\mathrm{x}^\om = c_\mathrm{x} \int \rho^{4/3}(\vec{r})
 f\left(\frac{p(\vec{r})}{\om}, s(\vec{r})\right) \de^3 \vec{r}
\end{equation}
with the new enhancement factor $f(\la,s)$ now being a function
of two dimensionless variables ($\la$ is length-like),
\begin{equation}
\label{eq:fq}
f(\la,s) = -\frct89 \int\limits_0^\infty r \erfc\left(\frac{r}{\la}\right) q(r,s) \de r .
\end{equation}
Finding a good shape function $q(r,s)$ given an enhancement factor $f(s)$
is the problem we want to solve here.
In doing so, we should respect the sign of $q(r,s) \le 0$, the normalization
\begin{equation}
\label{eq:s1}
\frct{4}{3\pi} \int\limits_0^\infty r^2 q(r,s) \de r = -1,
\end{equation}
and the energy connection
\begin{equation}
\label{eq:sf}
\frct89 \int\limits_0^\infty r\, q(r,s) \de r = -f(s),
\end{equation}
while the known on-top value and curvature~\cite{B83}
\begin{equation}
\label{eq:curv}
\lim_{r,s\to 0} q(r,s) = -\frct12 + \left(\frct1{10} - \frct1{27} s^2 \right) r^2 + \dots
\end{equation}
are very helpful to build a good overall shape.
The uniform electron gas has an oscillatory function
\begin{equation}
\label{eq:ueg}
\tilde{q}(r) = -\frct92 \left( \frac{\sin r - r\cos r}{r^3} \right)^2
\end{equation}
with a rather long tail of $-\frct94 r^{-4}$,
whereas finite band gap systems have it more localized and mostly smooth.
What we have written up to here is the common knowledge~\cite{PW92,PBW96,EP98} in the field,
with all this in mind, we will now build and compare the new models of our own.

We want Eq.~(\ref{eq:fq}) to have the exact uniform electron gas limit at $s = 0$,
and the only way to meet this is when
\begin{equation}
q(r,0) = \tilde{q}(r),
\end{equation}
so our first model will be an interpolation
\begin{equation}
\label{eq:q2w}
q(r,s) = \bigl(1 - w(s)\bigr) \tilde{q}(r) + w(s)\, q_2\bigl(r,a(s),c(s)\bigr)
\end{equation}
between $\tilde{q}(r)$ and a three-term Hermite function
\begin{eqnarray}
\nonumber
q_2(r,a,c) &=& -\Bigl[\frct12 + \left(2\sqrt{\pi}a^5 - \frct13 a^2 - \frct52 c \right) r^2 + c\, a^2 r^4 \Bigr]
\\ & & \times \exp\left(-a^2 r^2\right) .
\end{eqnarray}
It already follows Eq.~(\ref{eq:s1}), while from Eq.~(\ref{eq:sf}) we get
\begin{equation}
\label{eq:c}
c(s) = 4\sqrt{\pi} a^5(s) - \frct92 a^4(s) \left(1 + \frac{f(s) - 1}{w(s)}\right) + \frct13 a^2(s),
\end{equation}
and nothing seems to be more natural than
\begin{equation}
\label{eq:w}
w(s) = 1 - \exp\left(-\ga s^2 \right)
\end{equation}
with $\ga$ set as
\begin{equation}
\label{eq:g}
\ga = \frac{6075 \al^4 \mu - 20}{4320\sqrt{\pi}\al^5 - 6075\al^4 + 900\al^2 - 54}
\end{equation}
to fulfill Eq.~(\ref{eq:curv}), here $\al\equiv a(0)$
and $\mu$ is from
\begin{equation}
\label{eq:cf}
\lim_{s\to 0} f(s) = 1 + \mu s^2 + \dots
\end{equation}
(understanding that $f(s)$ should always bee~\cite{B88}
an \emph{even} function of $s$).
After all this, we are left with the freedom to choose a good function $a(s)$
limited mainly by the sign of $q(r,s) \le 0$.

The integral of Eq.~(\ref{eq:fq}) over the function of Eq.~(\ref{eq:q2w})
has a simple closed-form expression
\begin{equation}
\label{eq:f2}
f(\la,s) = \bigl(1 - w(s)\bigr) \tilde{u}(\la) + w(s) u_2\bigl(\la, a(s), c(s)\bigr)
\end{equation}
with the known~\cite{GAP96} uniform electron gas function
\begin{equation}
\label{eq:u0}
\tilde{u}(\la) = 1
 + \frac{2}{3\la^2}
 + \left(4 - \frac{2}{\la^2} \right)
 \frac{1 - \exp\left(-\la^2\right)}{3\la^2}
 - \frac{4\sqrt{\pi}\erf(\la)}{3\la}
\end{equation}
which for small $\la$ should be evaluated using (a few terms of) the series
\begin{equation}
\tilde{u}(\la) =
\sum\limits_{n=1}^\infty \frac{(-1)^{n-1}}{(n+2)!\, (n + \frct12)} \la^{2n} ,
\end{equation}
and the well-behaved functions
\begin{eqnarray}
\nonumber
u_2(\la, a, c) &=&
\frct49 \la^2 v_0 (a\la)
+ \frct89 \left( 2\sqrt{\pi} a^5 - \frct13 a^2 - \frct52 c \right)
 \la^4 v_1 (a\la) \\
&+& \frct89 c\, a^2 \la^6 v_2 (a\la) ,
\end{eqnarray}
\begin{equation}
\label{eq:vn}
\begin{array}{lcl}
v_n(t) &=& \upsilon_n\Bigl(\sqrt{1 + t^2} \Bigr), \\
\upsilon_0(x) &=& 1/\bigl(2x(1+x) \bigr) , \\
\upsilon_1(x) &=& \bigl(1 + 2x \bigr)/\bigl(4x^3 (1+x)^2 \bigr) , \\
\upsilon_2(x) &=& \bigl(3 + 9x + 8x^2 \bigr)/\bigl(8x^5 (1 + x)^3 \bigr).
\end{array}
\end{equation}

There are two kinds of enhancement factors:
either bounded by a constant, $1\le f(s) \le 1 + \ka$,
or unbounded $f(s)\to\infty$ as $s\to\infty$.
We will deal first with those of the former kind,
the simplest~\cite{B86} and widely used~\cite{PBE96}
\begin{equation}
\label{eq:fa}
f(s) = 1 + \ka - \ka/\left(1 + \mu s^2 /\ka \right) ,
\end{equation}
and another useful~\cite{HHN99}
\begin{equation}
\label{eq:fe}
f(s) = 1 + \ka - \ka\exp\left(-\mu s^2 /\ka \right) ,
\end{equation}
both having only two parameters derivable from first principles,
the gradient coefficient~\cite{MB68}
\begin{equation}
\label{eq:cm}
\mu\approx 0.219514972757702295113021595185752214898986
\end{equation}
(we have carefully computed the integrals~\cite{MB68} numerically
to all digits given),
and an estimated~\cite{LO81} $\ka = 0.804$
from the global lower bound~\cite{L79} on the exchange energy.

Our simplest $a(s)$ in Eqs.~(\ref{eq:q2w}) and~(\ref{eq:c}) is then a constant $a(s) = A$
whose value can be nailed down by setting
\begin{equation}
\label{eq:c0n}
\lim\limits_{s\to\infty}c(s) = 0,
\end{equation}
so that $A$ is a root of the cubic equation
\begin{equation}
24\sqrt{\pi} A^3 - 27 (1 + \ka) A^2 + 2 = 0,
\end{equation}
for $\ka = 0.804$ we get
\begin{equation}
A\approx1.106630987280122660996724171466275
,
\end{equation}
and for $\mu$ of Eq.~(\ref{eq:cm}), from Eq.~(\ref{eq:g}) with $\al=A$,
\begin{equation}
\label{eq:gn}
\ga\approx 0.426236616475280454832173010376694
.
\end{equation}
This is our simplest model that can also work with other more flexible~\cite{AB02}
forms of $f(s)$ as long as they are bounded by a constant,
it is straightforward to implement and it has, through Eq.~(\ref{eq:c}),
the input $f(s)$ as a multiplicative factor in the expression for $f(\la,s)$.
Plots show that $c(s) > 0$, monotonic for Eq.~(\ref{eq:fe})
but with a slight wave up and down for Eq.~(\ref{eq:fa}).

As a prototype of an unbounded $f(s)$, we take the most well-known and widely used~\cite{B88}
\begin{equation}
\label{eq:fb}
f(s) = 1 + \frac{\mu s^2}{1 + \mu\nu s \ln\bigl(\eta s + \sqrt{1 + \eta^2 s^2}\,\bigr)} ,
\end{equation}
\begin{eqnarray}
\nu &=& \sqrt[3]{9/(2\pi^2)} , \\
\label{eq:eta}
\eta &=& 2\sqrt[3]{6\pi^2} ,
\end{eqnarray}
where $\mu$ can be either adjusted~\cite{B88} to fit some data
or~\cite{C16} the theoretical constant of Eq.~(\ref{eq:cm}),
$\nu$ is fixed by the asymptotic behavior of the energy density,
(and we must note that $\eta$ could as well have been an adjustable parameter ---
its value of Eq.~(\ref{eq:eta}) is nothing but arbitrary).
Here, we should have an $a(s)$ that always grows with $s$,
otherwise there would have been $c(s) < 0$ and $q(r,s) > 0$.
To meet Eq.~(\ref{eq:c0n}), it can be shown that the first two terms of
\begin{equation}
\lim_{s\to\infty} a(s) = \frac{9f(s)}{8\sqrt{\pi}}
- \frac{16\sqrt{\pi}}{243 f^2(s)}
- \frac{4096\pi\sqrt{\pi}}{531441 f^5(s)}
+ \dots
\end{equation}
would have been needed, and the third and higher terms would help $c(s)$ reach zero faster.
We cannot take these first two terms exactly as written for $a(s)$, however,
because there would be $c(s) < 0$ for some small $s$, but the simplest
\begin{equation}
\label{eq:9f8pi}
a(s) = \frac{9f(s)}{8\sqrt{\pi}}
\end{equation}
already yields a working overall solution.

By the way, putting the bounded $f(s)$ of Eqs.~(\ref{eq:fa}) or~(\ref{eq:fe})
into Eq.~(\ref{eq:9f8pi}) would also work and give us another $f(\la,s)$ of Eq.~(\ref{eq:f2})
that is clearly not the same as our first model with the constant $a(s)=A$,
and when we plot the ratio of these $f(\la,s)$, we see that the one based on Eq.~(\ref{eq:9f8pi})
is down to $15\%$ smaller for some $\la$ and $s$.
This gives us a hint at their diversity and makes us think of how to narrow down
the choice of $a(s)$.
Besides Eq.~(\ref{eq:c0n}), we can nail it down at the other end, $s=0$, by
\begin{equation}
c(0) = 0,
\end{equation}
which makes $\al\equiv a(0)$ the root of the seventh-degree polynomial parametrized by $\mu$,
that can be written as
\begin{equation}
\label{eq:am}
\frac{240\sqrt{\pi} \al^3 - 270 \al^2 + 20}{14580\sqrt{\pi} \al^7 - 6075 \al^4 + 729 \al^2} = \mu ,
\end{equation}
for $\al$ has to be solved for together with $\ga$ of Eq.~(\ref{eq:g}).
For $\mu$ of Eq.~(\ref{eq:cm}) we get
\begin{equation}
\begin{array}{lcl}
\al &\approx& 0.535666481894751540210390580247207
, \\
\ga &\approx& 2.149928276780137518154670987133133
,
\end{array}
\end{equation}
and this $\ga$ is roughly $5$ times greater than that of Eq.~(\ref{eq:gn}),
we think the greater $\ga$ to be better because then the oscillatory $\tilde{q}(r)$
fades away more quickly in Eq.~(\ref{eq:q2w}).
It is easy to build monotonic interpolations $\al \le a(s) \le A$
for a bounded $f(s)$ to get a small $c(s)>0$: for Eq.~(\ref{eq:fa})
\begin{eqnarray}
\label{eq:aa}
a(s) &=& \al + a_2 s^2 /\left(1 + a_2 s^2 /(A - \al) \right), \\
a_2 &=& (A - \al)^2 \left(9(1 + \ka) A^2 - 2 \right) \mu \left/ \left(9\ka^2 A^3 \right) \right. ,
\end{eqnarray}
yields $c(s) < 0.023$ for all $s$; whereas for Eq.~(\ref{eq:fe})
\begin{eqnarray}
\label{eq:ae}
\nonumber
a(s) &=& A - a_\mathrm{e} \exp\left(-\mu s^2 /\ka \right) \\
  &+& (a_\mathrm{e} + \al - A) \exp\left(-2\mu s^2 /\ka \right) , \\
a_\mathrm{e} &=& 9\ka A^3 /\left(9(1+\ka) A^2 - 2\right) ,
\end{eqnarray}
yields $c(s) < 0.016$ for all $s$.
Likewise, for Eq.~(\ref{eq:fb})
\begin{eqnarray}
\label{eq:ab}
a(s) &=&
\frct{9}{8\sqrt{\pi}}\cdot\frac{1 + \mu\nu s \asinh(\eta s) + \mu s^2}{1 + \mu\nu s \asinh(\eta s)}
\\ \nonumber
&+&
\frac{\al - \frct{9}{8\sqrt{\pi}}
 + \frct{16\sqrt{\pi}}{243} - \frct{16\sqrt{\pi}}{243} \bigl(1 + \mu\nu s \asinh(\eta s)\bigr)^2}
 {\bigl(1 + \mu\nu s \asinh(\eta s) + \mu s^2 \bigr)^2}
\end{eqnarray}
makes $c(s) < 0.036$ for all $s$.
This experience helps us get rid of $c(s)$ altogether,
ending up with an even simpler model
\begin{eqnarray}
\label{eq:q1w}
q(r,s) &=& \bigl(1 - w(s)\bigr) \tilde{q}(r) + w(s) q_1\bigl(r,a(s)\bigr) , \\ \nonumber
q_1(r,a) &=& -\Bigl[\frct12 + \left( 2\sqrt{\pi} a^3 - \frct13 \right) a^2 r^2 \Bigr]
 \exp\left(-a^2 r^2 \right) ,
\end{eqnarray}
\begin{eqnarray}
\label{eq:f1}
f(\la,s) &=& \bigl(1 - w(s)\bigr) \tilde{u}(\la) + w(s) u_1\bigl(\la, a(s)\bigr) , \\ \nonumber
u_1(\la, a) &=&
\frct49 \la^2 v_0 (a\la)
+ \frct89 \left( 2\sqrt{\pi} a^5 - \frct13 a^2 \right)
 \la^4 v_1 (a\la) ,
\end{eqnarray}
that can be used in two ways:
either by redefining
\begin{equation}
w(s) = \frac{27 a^2(s) \bigl(f(s) - 1\bigr)}{24\sqrt{\pi} a^3(s) - 27 a^2(s) + 2}
\end{equation}
for use with some $a(s)$ like in Eqs.~(\ref{eq:aa}), (\ref{eq:ae}), and (\ref{eq:ab});
or by holding true to Eq.~(\ref{eq:w})
while fearlessly solving the cubic equation for $a(s)$ to get
\begin{equation}
\label{eq:af}
a(s) = \left[ 3\sqrt{2}\,\varphi(s) \sin \left(\frac13 \asin \frac{4\sqrt{2\pi}}{9\varphi^3(s)}\right) \right]^{-1} ,
\end{equation}
\begin{equation}
\label{eq:fw}
\varphi(s) = \sqrt{1 + \bigl(f(s) - 1 \bigr)/w(s)} .
\end{equation}
This last idea is so strikingly simple that nothing is left to be shaved away with Ockham's razor,
and we like it the most.

Thus, given a $f(s)$ of any meaningful kind,
we find its $\mu$ of Eq.~(\ref{eq:cf}), get $\al$ from Eq.~(\ref{eq:am}),
and $\ga$ from Eq.~(\ref{eq:g}), so we have $w(s)$ of Eq.~(\ref{eq:w}),
hence $a(s)$ of Eqs.~(\ref{eq:af}) and~(\ref{eq:fw}),
that yields us $f(\la,s)$ of Eq.~(\ref{eq:f1}) with Eqs.~(\ref{eq:u0}) and~(\ref{eq:vn}).

Here our tale would have had a happy end,
but we feel that someone may call it a~\textit{heresy}
to work with an oscillatory shape function having a thin but too long tail.
In the spirit of the early works~\cite{PW92,EP98,HJS08},
we will now build some new and simple non-oscillatory models
(of interest on their own) and compare the outcomes $f(\la,s)$ one-to-one
to see only a small difference.

We begin with our two new amazingly beautiful non-oscillatory exchange hole models
for the uniform electron gas
that follow Eqs.~(\ref{eq:s1}), (\ref{eq:sf}), (\ref{eq:curv})
and have the $-\frct94 r^{-4}$ tail from Eq.~(\ref{eq:ueg}):
the split-exponent version
\begin{eqnarray}
 \nonumber
\bar{\bar{q}}(r) &=&
-\frct94 \left[1 - \left(1 + \be^2 r^2 \right) \exp\left(-\be^2 r^2\right)\right] r^{-4}
\\ \nonumber
&-& \Bigl[\frct12 - \frct98 \be^4
+ \left(\frct12 \al^2 - \frct98 \al^2 \be^4 + \frct34 \be^6 - \frct1{10}\right) r^2
 \Bigr]
\\
&& \times \exp\left(-\al^2 r^2 \right) ,
\label{eq:qab}
\end{eqnarray}
$\al$ and $\be$ being roots of the polynomial system
\begin{equation}
\left\{\begin{array}{rcr}
90 \be^6 + \left( 100 - 225 \be^4 \right) \al^2 + \bigl(360 \be - 240\sqrt{\pi} \bigr)\al^5 &=& 12, \\
\left(45 \be^2 - 45\right) \al^4 - \left(45 \be^4 - 20 \right) \al^2 + 15 \be^6 &=& 2,
\end{array}\right.
\end{equation}
\begin{equation}
\begin{array}{l}
\be\approx 1.018374108606862926108404234418115726
, \\
\al\approx 0.731832253115642144951075675174173985
,
\end{array}
\end{equation}
and the shared-exponent version
\begin{eqnarray}
\label{eq:qa}
\nonumber
\bar{q}(r) &=&
-\frct94 \left[1 - \left(1 + \al^2 r^2 \right)\exp\left(-\al^2 r^2\right)\right] r^{-4}
\\
&-& \Bigl[\frct12 - \frct98 \al^4
 + \bigl(\frct12 \al^2 - \frct38 \al^6  - \frct1{10}\bigr) r^2
\\ \nonumber
&+& \bigl(\frct1{25} \al^2 - \frct13 \al^4 + \frct{4\sqrt{\pi}}{5} \al^7 - \frct34 \al^8 \bigr) r^4
\Bigr] \exp\left(-\al^2 r^2 \right) ,
\end{eqnarray}
\begin{equation}
\label{eq:a0}
225 \al^6 - 480\sqrt{\pi} \al^5 + 675 \al^4 - 100 \al^2 + 6 = 0 ,
\end{equation}
\begin{equation}
\label{eq:a1}
\al\approx 0.780787542771215180413790059432608235
.
\end{equation}
In both cases, all three functions $q(r)$, $rq(r)$, and $r^2 q(r)$ have slim shapes without any shoulders
for $r \ge 0$,
whereas their forerunners~\cite{EP98,HJS08}, to become shoulderless, needed one more degree of freedom
to be fixed by a sophisticated information-entropy-maximization~\cite{J57} principle.
We hope that our finding may help others in their future work.

From Eq.~(\ref{eq:qa}), we build
\begin{eqnarray}
\label{eq:q2a}
\nonumber
\bar{q}(r,s)
&=& -\frct94 \left[1 - \left(1 + \al^2 r^2 \right)\exp\left(-\al^2 r^2\right)\right] r^{-4} \\ \nonumber
&& \times\exp\left(-h^2(\chi s)\, a^2(s)\, r^2 - \ga s^2 \right) \\ \nonumber
&-& \left[ \frct12 - \frct98 \al^4 \exp\left(-\ga s^2 \right) + b(s) r^2 + c(s) r^4 \right] \\
&& \times\exp\left(-a^2(s)\, r^2 \right)
\end{eqnarray}
with $a(0)=\al$ of Eq.~(\ref{eq:a0}),
while $b(s)$ and $c(s)$ have to fulfill Eqs.~(\ref{eq:s1}) and~(\ref{eq:sf}),
\begin{eqnarray}
\nonumber
b(s) &=& \left(12 z(s) - 8\sqrt{\pi}\right) a^5(s)
+ \frct{45}4 \bigl(f(s) - y(s)\bigr) a^4(s) \\
&+& \left(\frct{21}8 \al^4 \exp(-\ga s^2) - \frct76 \right) a^2(s) ,
\\ \nonumber
c(s) &=& \left(4 \sqrt{\pi} - 6 z(s)\right) a^7(s)
+ \frct92 \bigl(y(s) - f(s)\bigr) a^6(s) \\
&+& \left(\frct13 - \frct34 \al^4 \exp(-\ga s^2) \right) a^4(s) ,
\end{eqnarray}
using the integrals over the first $r^{-4}$ term of Eq~(\ref{eq:q2a}),
\begin{eqnarray}
z(s) &=& \al Z\left(\al^{-1} h(\chi s) a(s) \right) \exp\left(-\ga s^2 \right) , \\
\label{eq:ys}
y(s) &=& \al^2 Y\left(\al^{-1} h(\chi s) a(s) \right) \exp\left(-\ga s^2 \right) ,
\end{eqnarray}
\begin{eqnarray}
Z(x) &=& \frac{1 + 2x^2}{\sqrt{1 + x^2}} - 2x , \\
\label{eq:yx}
Y(x) &=& 1 + x^2 \ln\frac{x^2}{1 + x^2} ,
\end{eqnarray}
for $x \gg 1$, these should be computed as
\begin{eqnarray}
Z(x) &=& \frac{1}{\Bigl(1 + 2x^2 + 2x\sqrt{1 + x^2}\Bigr)\sqrt{1 + x^2}} , \\
Y(x) &=& 1 - x^2 \ln\left(1 + x^{-2}\right)
= \sum\limits_{k=0}^\infty \frac{(-1)^k}{(k+2) x^{2+2k}} .
\end{eqnarray}
To follow the curvature of Eq.~(\ref{eq:curv}) at $s=0$, we need
\begin{equation}
\label{eq:g1}
\ga = \frac{27 \al^4 - 48\sqrt{\pi} \al^3 + 54 \al^2 - 4}{9 \al^5} a''(0)
+ \frac{3}{\al^2} \mu - \frac{4}{405 \al^6} .
\end{equation}
The first term in Eq.~(\ref{eq:q2a}) should smoothly switch from having the $-\frct94 r^{-4}$ tail
to a short-range exponential behavior,
to overcome the logarithmic singularity as $s\to 0$ in its integrals over $r$,
such as in Eqs.~(\ref{eq:ys}) and~(\ref{eq:yx}),
we multiply $a(s)$ by the healing function $h(\chi s)$,
\begin{equation}
h(x) = \exp\left(-\frac{\exp\left(\frct12 - x^2\right)}{\sqrt{2}\,|x|}\right) ,
\end{equation}
that has its value and all derivatives zero at $s=0$.
For $\chi$, we can take either the greatest $\chi_0$ that still yields
$b(s) \ge 0$, or the greatest $\chi_1$ that still yields a monotonic $b(s)$, by solving
\begin{equation}
\label{eq:ch}
\left\{\begin{array}{rcl}
b^{(n)} (s_n; \chi_n) &=& 0, \\
b^{(n+1)} (s_n; \chi_n) &=& 0,
\end{array}\right.
\end{equation}
for $s_n$ and $\chi_n$, where $b^{(n)}(s;\chi) \equiv \partial^n b(s;\chi)/\partial s^n$;
we can also set $\chi = 0$ to see what happens.
Given some $a(s)$, there is an explicit closed-form expression for Eq.~(\ref{eq:fq}),
\begin{eqnarray}
\label{eq:fh}
\nonumber
\bar{f}(\la,s)
&=& \bar{u}\bigl(\la,\al, h(\chi s)\, a(s)\bigr) \exp\left(-\ga s^2 \right)
\\
&+& \Bigl(\frct49 - \al^4 \exp\left(-\ga s^2 \right)\Bigr) \la^2
 v_0 \bigl(a(s)\la \bigr)
\\ \nonumber
&+& \frct89 b(s) \la^4 v_1 \bigl(a(s) \la\bigr)
 +  \frct89 c(s) \la^6 v_2 \bigl(a(s) \la\bigr) ,
\end{eqnarray}
\begin{eqnarray}
\nonumber
\bar{u}(\la,\al,\be)
&=& \al^2      \bar{v}  \left((\al^2 + \be^2) \la^2, \be^2 \la^2\right) \\
&+& \be^2 \bar{\bar{v}} \left((\al^2 + \be^2) \la^2, \be^2 \la^2\right) ,
\end{eqnarray}
\begin{eqnarray}
\bar{v}(\sigma,\tau) &=&
\frac{\displaystyle \frac{\sigma}{1 + \sqrt{1 + \sigma}} + \frac{\tau}{1 + \sqrt{1 + \tau}}}{\sqrt{1 + \sigma} + \sqrt{1 + \tau}} ,
\\
\bar{\bar{v}}(\sigma,\tau) &=&
2\ln\frac{1 + \sqrt{1 + \tau}}{1 + \sqrt{1 + \sigma}} .
\end{eqnarray}

\begin{figure}
\includegraphics[scale=1.0]{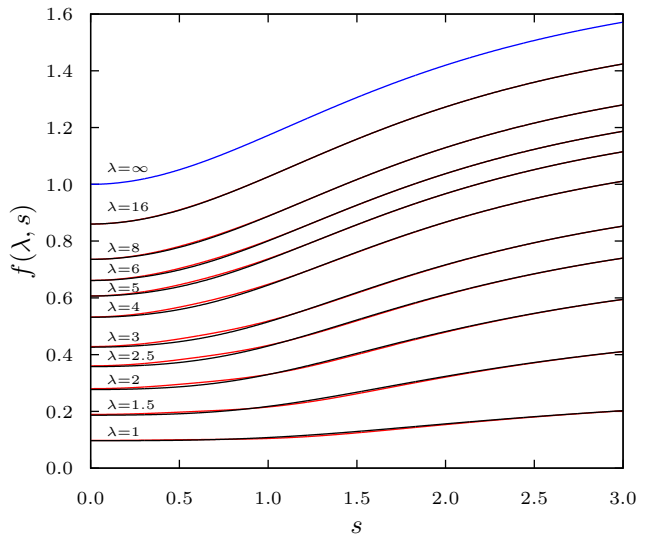}
\caption{\label{fig:1} Comparison of enhancement factors,\\
black: $f(\la,s)$ of Eqs.~(\ref{eq:f1}) and~(\ref{eq:af}), \\
red: $\bar{f}(\la,s)$ of Eq.~(\ref{eq:fh}) with $\chi = \chi_1$.}
\end{figure}

\begin{figure}
\includegraphics[scale=1.0]{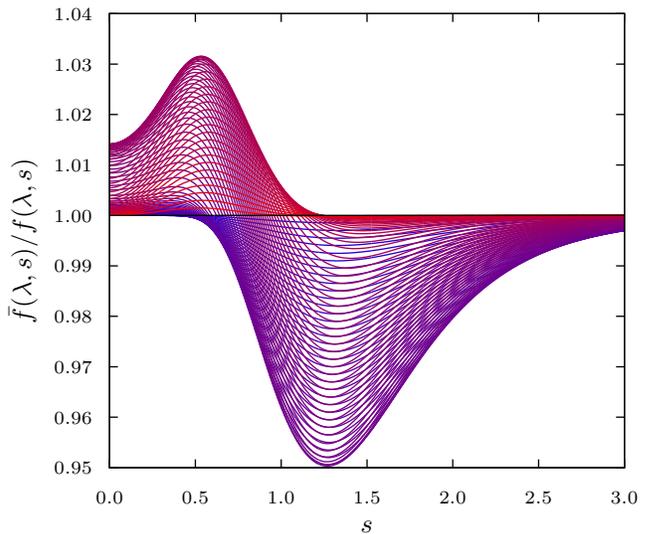}
\caption{\label{fig:2} Comparison of enhancement factors:\\
the ratio $\bar{f}(\la,s)/f(\la,s)$ as a function of $s$ and $\la$.}
\end{figure}

For $f(s)$ of Eq.~(\ref{eq:fa}),
we take $a(s)$ from Eq.~(\ref{eq:aa}) with $\al$ from Eq.~(\ref{eq:a1}), $a''(0) = 2a_2$ for $\ga$ of Eq.~(\ref{eq:g1}),
solve Eq.~(\ref{eq:ch}) to get
\begin{equation}
\begin{array}{l}
\chi_0 \approx 0.489520810866673031024609627137989575
, \\
\chi_1 \approx 0.346299740672885863135065914142974003
,
\end{array}
\end{equation}
and now we can compare $\bar{f}(\la,s)$ of Eq.~(\ref{eq:fh})
to our best $f(\la,s)$ of Eqs.~(\ref{eq:f1}) and~(\ref{eq:af}).
In Fig.~\ref{fig:1} the two functions are plotted
and we see how regular they are and how little they differ,
even more impressive is the colorful family of curves in Fig.~\ref{fig:2}
for their ratio $\bar{f}(\la,s)/f(\la,s)$.
This way, we get a measure of their similarity,
\begin{equation}
\begin{array}{lclclcl}
             &       & 1     &\le& \bar{f}(\la,0)/\tilde{u}(\la) &<& 1.0142, \\
\chi=0:      &\qquad & 0.986 &<&   \bar{f}(\la,s)/f(\la,s)       &<& 1.042,  \\
\chi=\chi_1: &\qquad & 0.950 &<&   \bar{f}(\la,s)/f(\la,s)       &<& 1.032,  \\
\chi=\chi_0: &\qquad & 0.903 &<&   \bar{f}(\la,s)/f(\la,s)       &<& 1.023.  \\
\end{array}
\end{equation}
In the uniform electron gas limit,
the nodeless shape function of Eq.~(\ref{eq:qa}) yields
the integral $\bar{f}(\la,0)$
that matches the exact $\tilde{u}(\la)$ of Eq.~(\ref{eq:u0})
to within $1.5\%$, while for $s>0$
the models are only a few times farther away from each other,
being closest for $\chi=0$.
Thus, $\chi$ plays no dramatic role, and it seems better to cut the tail 
in depth by $\exp\left(-\ga s^2 \right)$
than at length by $\exp\left(-h^2(\chi s)\, a^2(s)\, r^2\right)$,
to enjoy a rewarding simplification of the equations.
In this way, the function of Eq.~(\ref{eq:qab}) can also be used,
and a cubic equation for $a(s)$ can then be set up,
but we leave it out here to save space.

It is now clear that both kinds of shape functions ---
both the oscillatory of Eq.~(\ref{eq:q1w}) and
the non-oscillatory of Eq.~(\ref{eq:q2a}) ---
would yield nearly the same integral output
of Eq.~(\ref{eq:fq}) under the same constraints of Eqs.~(\ref{eq:s1}),
(\ref{eq:sf}), and~(\ref{eq:curv}).
To our mind, the oscillatory function gives the best solution:
we get an explicit closed-form expression for $f(\la,s)$
in terms of the given $f(s)$
using Eqs.~(\ref{eq:w}), (\ref{eq:af}), (\ref{eq:fw}), and~(\ref{eq:f1});
furthermore, it has the exact uniform electron gas limit.
Nevertheless, our experience with the non-oscillatory functions
was not in vain and these can be used in the further work
on new functionals.

It might be time for a thorough benchmark of the new model
on a wide set of molecules, but we put it off for now
until we learn how to combine it
with a dispersion-correction functional~\cite{DRSLL04,VV10}.

We find our long-range corrected version of the PBE~\cite{PBE96} functional
with $\frac1\om = 3\;\mathrm{au}$ (an easy-to-remember whole number)
to be already a good next step after its $\frct14$-fixed-fraction hybrid~\cite{PEB96,AB99},
and it can be used routinely in mechanistic studies
of molecular structure and reactivity
toward a full understanding of chemical kinetics.

\clearpage

\end{document}